\documentstyle[aps,graphics,amsmath,12pt,preprint]{revtex}\def\wideabs{}

\tightenlines
\begin{document}
\wideabs{
\title{Adsorption of a fluid in an aerogel: integral equation 
approach}
\author{V. Krakoviack\thanks{Present address: Department of Chemistry,
University of Cambridge, Lensfield Road, Cambridge CB2 1EW, United
Kingdom}, E. Kierlik, M.-L. Rosinberg and G. Tarjus}
\address{Laboratoire de Physique Th{\'e}orique des Liquides,
Universit{\'e} Pierre et Marie Curie, \\
4 Place Jussieu, 75252 Paris Cedex 05, France}
\date{\today}
\maketitle

\begin{abstract}
We present a theoretical study of the phase diagram and the structure
of a fluid adsorbed in high-porosity aerogels by means of an
integral-equation approach combined with the replica formalism. To
simulate a realistic gel environment, we use an aerogel structure
factor obtained from an off-lattice diffusion-limited
cluster-cluster aggregation process. The predictions of the theory are
in qualitative agreement with the experimental results, showing a
substantial narrowing of the gas-liquid coexistence curve (compared to
that of the bulk fluid), associated with weak changes in the critical
density and temperature. The influence of the aerogel structure
(nontrivial short-range correlations due to connectedness,
long-range fractal behavior of the silica strands) is shown to be
important at low fluid densities. 
\end{abstract}
}

\section{Introduction}

The thermodynamics of fluids imbibed in disordered porous materials is
a subject of great interest, both from an experimental and from a
theoretical perspective \cite{FLC1994}. Of special importance 
is the question of fluid-fluid phase transitions which are found
experimentally to be drastically altered compared to the bulk
situation. One of these transitions is the so-called capillary
condensation observed during the adsorption of a single-component
gas  at low temperature. It is marked by a  steep
increase in the adsorbed mass that usually occurs  at some pressure below the saturated vapor
pressure of the bulk gas.

From an experimental point of view, one has to distinguish two types
of systems. A first class of adsorbents consists of mesoporous
materials of low porosity (between 30 and 60\%), like porous glasses and
silica gels. Inside such a matrix, capillary condensation is usually
accompanied by a pronounced hysteresis loop that shrinks and
eventually disappears at some temperature $T_h$ less than $T_c$, the
bulk critical temperature. This  (reproducible) hysteresis
loop prevents the direct observation of a thermodynamic, equilibrium phase
transition,  and  the mere existence of a true phase transition
underlying capillary condensation in these systems remains a
controversial issue. The situation appears clearer  for a second class
of adsorbents, namely very dilute silica networks like aerogels with
porosities as high as 95-98\% \cite{FFC1991,WC1990,Z1996}. There is
indeed experimental evidence that fluids in such media undergo
true phase transitions that terminate at a genuine critical point.
 Despite the very low density of the
adsorbing solids, these transitions display features strikingly
different from those of bulk fluids. Thus, the phase diagrams of
${}^4$He and N$_2$ in aerogels show a marked narrowing of the
liquid-vapor coexistence curve and a
displacement of the critical point to a lower temperature and a higher
density  compared to that in the bulk \cite{WC1990}. A similar behavior has been observed for a
mixture of isobutyric acid and water in a very dilute silica gel
\cite{Z1996}.

Recently, the study of a disordered lattice-gas model by means of
mean-field density functional theory has provided new insight into
the physical origin of the phenomenology associated with capillary condensation\cite{KRTV2001}. First, it
has been found that true equilibrium phase transitions may occur only when
the perturbation induced by the solid is sufficiently small. It is
thus likely that true phase transitions may occur in aerogels, but
not  in dense matrices of low porosity. Secondly, it
has been shown that hysteresis can occur with or without an underlying
phase transition and that, if a transition exists, the disappearance of
the hysteresis loop is not associated to capillary criticality. This
results from the fact that there is a range of temperature and
chemical potential for which the system has many metastable
states ; it relies on the assumption, expected to be valid for
low-porosity solids, but not necessarily for very
dilute gels, that
the fluid is not able to fully equilibrate after a change in chemical potential
(or pressure). 

In the last few years, the concept of ``quenched-annealed'' binary
mixture has been widely used to study the structure
 and the phase behavior of
fluids adsorbed in disordered porous matrices. In this approach,
pioneered by Madden and Glandt \cite{MG1988} and further developed
by the application of the replica method 
\cite{G1992,RTS1994}, the fluid molecules equilibrate
in a matrix of particles frozen in a disordered configuration that is
sampled from a given probability distribution. On this basis, computer
simulations and
integral-equation  theories  using extensions of standard liquid-state
approximations have been proposed\cite{R1999}. But when 
the physics of adsorption is  dominated by the existence of
many metastable states (as shown in Ref. \cite{KRTV2001}), it is 
expected that these methods, that have proved very efficient
in dealing with bulk fluids and fluids in simple external potentials,
could run into difficulties and overestimate the possibility of true 
phase transitions.

The case of very dilute silica gels is the most likely exception in which 
these approaches could remain fully operational.  Accordingly, we
report in the present paper an investigation
of the phase behavior of a simple fluid adsorbed in an aerogel
within the framework of the integral-equation theory of
quenched-annealed binary mixtures. This study is based on the 
optimized cluster theory (OCT) approach\cite{AC1972} whose application to
quenched-annealed mixtures was first considered in  Ref.\cite{KRTM1997}. It can be considered as an extension of
previous work toward a description of the actual
experimental systems, since, at variance with previous studies of
quenched-annealed mixtures, a more realistic model of the solid is
used. Indeed, aerogels display very specific structural
properties that may play an important role for fluid adsorption.
In the past,  this has motivated  the development of
various lattice models taking into account all or part of the specific
features of the aerogel-induced disorder ;  the phase
transitions of these models have been studied by various methods, both
theoretically and by computer simulation
\cite{M1991,SP1992,FB1995,UHJ1995,STC1999}. The present
work relies on the off-lattice version \cite{HFPJ93} of the model used
in Refs. \cite{UHJ1995,STC1999}, where the aerogel was described  as a
connected cluster of sites generated by a diffusion limited
cluster-cluster aggregation (DLCA) process \cite{M1983,KBJ1983}.

In the next section, the theoretical framework used in this paper is
summarized. In section \ref{sec-model}, the molecular model for the fluid
and the matrix are presented, and different aspects of the calculation
are discussed. The results are given  and discussed in section
\ref{sec-results}. 

\section{Theory}
\label{sec-theory}

In this section, we first summarize the theoretical expressions on
which the present calculation is based. The reader is referred to the
original papers for details\cite{KRTM1997}. We then present and
discuss the extension of the formalism needed  to treat the problem at hand. 

The replica-symmetric Ornstein-Zernike (RSOZ)  equations relating the
total pair correlation functions  $h_{\alpha \beta}(r)$ of a
quenched-annealed mixture to the corresponding direct correlation
functions $c_{\alpha \beta}(r)$
have been first obtained by Given and Stell \cite{G1992} and read:    
\begin{subequations}
\begin{align}
&h_{00}=c_{00}+\rho_0\,c_{00}\otimes h_{00}, \label{roz1}\\
&h_{10}=c_{10}+\rho_0\,c_{10}\otimes h_{00}+\rho_1\,c_c\otimes h_{10}, \\
&h_{11}=c_{11}+\rho_0\,c_{10}\otimes h_{01}+\rho_1\,c_c\otimes h_{11}
+\rho_1\,c_b \otimes h_c,\\
&h_{c}=c_{c}+\rho_1\,c_c\otimes h_c,
\end{align}
\end{subequations}
where the indices 0 and 1 refer to the matrix and fluid particles
respectively, $\otimes $ denotes a convolution, and the indices $c$ and $b$ denote the so-called connected
and blocking (or disconnected) parts of $h_{11}$ and $c_{11}$ that obey
\begin{align}
h_{11}=h_b+h_c,\\
c_{11}=c_b+c_c.
\end{align}

The replica OCT \cite{KRTM1997} is applicable to systems with pairwise
additive 
interaction potentials and relies on the separation of these
potentials $w_{\alpha\beta}(r)$ into a (repulsive) part
$u^R_{\alpha\beta}(r)$ and a perturbative contribution
$u_{\alpha\beta}(r)$. The properties of the so-called reference
system, in which particles interact through the reference potentials,
are assumed to be known, and the central quantities of the theory are
then  renormalized potentials denoted as ${\cal C}_{\alpha\beta}(r)$.
Following Ref.\cite{KRTM1997}, we quote here the equations obtained when
only hard-sphere reference potentials with diameters
$\sigma_{\alpha\beta}$ are considered and when the  matrix particles interact with
each other only through such a hard-core interaction (i.e., $w_{00}(r)=u^R_{00}(r))$.

The renormalized potentials can be defined through the replica
optimized random-phase approximation (ORPA)
closure which reads ($\beta=1/(k_BT) $ where $k_B$ is   Boltzmann's constant
and $T$ is  the temperature)
\begin{subequations}
\begin{align}
&c_{00}(r)=c_{00}^R(r)\quad \mbox{for} \quad r > \sigma_{00},\label{disc1}\\
&c_{10}(r)=c_{10}^R(r)-\beta\,u_{10}(r)\quad \mbox{for} \quad r >\sigma_{10},\\
&c_{11}(r)=c_{11}^R(r)-\beta\,u_{11}(r)\quad \mbox{for} \quad r >\sigma_{11},\\
&c_b(r)=c_b^R(r),
\end{align}
\end{subequations}
complemented with the core conditions for the pair distribution
functions 
\begin{subequations}
\begin{align}
g_{00}(r)=0 \quad \mbox{for} \quad r < \sigma_{00},\label{disc2}\\
g_{10}(r)=0 \quad \mbox{for} \quad r < \sigma_{10},\\
g_{11}(r)=0 \quad \mbox{for} \quad r < \sigma_{11},
\end{align}
\end{subequations}
(there is no core condition for $g_b(r)=h_b(r)+1$).
Then, one has for the renormalized potentials:
\begin{subequations}
\begin{align}
&{\cal C}_{10}(r)=h_{10}(r)-h_{10}^R(r),\\
&{\cal C}_{11}(r)=h_{11}(r)-h_{11}^R(r),\\
&{\cal C}_{b}(r)=h_{b}(r)-h_{b}^R(r).
\end{align}
\end{subequations}

The ORPA free energy of the quenched-annealed system can be determined
in a closed form as
\begin{equation}\begin{split}
\overline{\cal{A}}_{\text{ORPA}}&=-\frac{\beta\,
\overline{A}_{\text{ex}}}{V}=\\
\overline{\cal{A}}^R&+\frac{1}{2}[\rho_1^2(\hat{c}_c(0)-\hat{c}_c^R(0))
+2\rho_1\rho_0(\hat{c}_{10}(0)-\hat{c}_{10}^R(0))] 
-\frac{\rho_1}{2} [c_{11}(0)-c_{11}^R(0)]\\
&-\frac{1}{2 (2 \pi^3)} \int d{\bf k} \{\ln [1-\rho_1\,\hat{c}_c(k)]-
\frac{\rho_1}{1-\rho_1\,\hat{c}_c(k)}[\hat{c}_b^R
(k)+\rho_0\frac{\hat{c}_{01}(k)^2}{1-\rho_0\,\hat{c}_{00}(k)}]\}\\
&+\frac{1}{2 (2 \pi^3)} \int d{\bf k} \{\ln [1-\rho_1\,\hat{c}_c^R(k)]-
\frac{\rho_1}{1-\rho_1\,\hat{c}_c^R(k)}[\hat{c}_b^R(k)
+\rho_0\frac{\hat{c}_{01}^R(k)^2}{1-\rho_0\,\hat{c}_{00}(k)}]\},
\end{split}\end{equation}
or, equivalently, by using the RSOZ equations,
\begin{equation}\begin{split}
\overline{\cal{A}}_{\text{ORPA}}&=\\
\overline{\cal{A}}^R&+\frac{1}{2}[\rho_1^2(\hat{c}_c(0)-
\hat{c}_c^R(0))+\rho_1\rho_0(\hat{c}_{10}(0)-\hat{c}_{10}^R(0))]
-\frac{\rho_1}{2} [c_{11}(0)-c_{11}^R(0)]\\
&-\frac{1}{2 (2 \pi^3)} \int d{\bf k} \ln \frac{1-
\rho_1\,\hat{c}_c(k)}{1-\rho_1\,\hat{c}_c^R(k)}-\beta\frac{\rho_1\,\rho_0}{2}
\int d{\bf r}\,g_{01}(r)\,u_{01}(r)\\
&+\frac{\rho_1^2}{2}\int d{\bf r}[h_c(r)-h_c^R(r)]c_b^R(r) + 
\frac{\rho_1\,\rho_0}{2}\int d{\bf r}[h_{01}(r)-h_{01}^R(r)]c_{01}^R(r),
\end{split}\end{equation}
where $\overline{\cal{A}}^R$ is the free energy of the reference system.

This ORPA free energy corresponds to the so-called energy route, since it
satisfies the Gibbs-Helmholtz equation
\begin{equation}
\overline{U}_{\text{ex}}=\left.
\frac{\partial\beta\overline{A}_{\text{ex}}}%
{\partial \beta}\right|_{V,\rho_0},
\end{equation}
where $\overline{U}_{ex}$ is the configurational internal energy of
the fluid inside the matrix and is given by 
\begin{equation}
\frac{\overline{U}_{ex}}{V}=\frac{1}{2} \int d{\bf
r}[\rho_1^2\,g_{11}(r)\, u_{11}(r)+2\rho_1\,\rho_0\,g_{01}(r)\,u_{01}(r)].
\end{equation}

The mean-spherical approximation (MSA)  is obtained from the ORPA when
the reference system is described by the Percus-Yevick (PY)
approximation, which corresponds to
\begin{subequations}
\begin{align}
&c_{00}^R(r)=0 \quad \mbox{for} \quad r > \sigma_{00},\label{disc3}\\
&c_{10}^R(r)=0 \quad \mbox{for} \quad r > \sigma_{10},\\
&c_{11}^R(r)=0 \quad \mbox{for} \quad r > \sigma_{11},\\
&c_b^R(r)=0 \quad \mbox{for} \quad r > 0.
\end{align}
\end{subequations}
Within this latter approximation, an expression for the excess chemical
potential can be derived:
\begin{equation}\begin{split}
\beta [\mu_{ex}-\mu_{ex}^R]&=\frac{1}{2}[c_{11}(0)-c_{11}^R(0)]-
\rho_1[\hat{c}_{11}(0)-\hat{c}_{11}^R(0)]\\
&-\rho_0[\hat{c}_{01}(0)-\hat{c}_{01}^R(0)]\\
&=\frac{\beta}{2}\int d{\bf r} [\rho_1\,g_{11}(r)\,u_{11}(r)+ 
\rho_0\,g_{01}(r)\,u_{01}(r)]\\
&-\frac{\rho_1}{2}[\hat{c}_{11}(0)-\hat{c}_{11}^R(0)]-
\frac{\rho_0}{2}[\hat{c}_{01}(0)-\hat{c}_{01}^R(0)].
\end{split}\end{equation}

As is often the case with simple closures, both the ORPA and the MSA
are thermodynamically inconsistent, which means that the free energy
obtained from the above energy route differs from that obtained by
integration with respect to the fluid density of the compressibility equation \cite{RTS1994,FG1994b} 
\begin{equation}\label{compress}
\frac{\beta}{\rho_1 \chi}=1-\rho_1\,\hat{c}_c(0),
\end{equation}
 where $\chi$ is the isothermal compressibility of the fluid inside the
matrix. This is an important remark, since it turns out that the
compressibility route does not yield a critical point in 3 dimensions\cite{PRST1995},
whereas the energy route provides one, albeit with an associated classical (mean-field)
behavior.

Finally, by truncating the OCT diagrammatic expansion to the first order
beyond ORPA, one obtains the ORPA+B2 approximation for the free
energy\cite{KRTM1997},  that is the sum of the ORPA result, Eq. (9),
and of the corrective term
\begin{equation}\begin{split}
B_2&=
\frac{1}{2}\rho_1\,\rho_0\int d{\bf r}\,h^R_{01}(r)
[{\cal C}_{01}(r)]^2+\rho_1\,\rho_0 \int d{\bf r}\,g^R_{01}(r)
\sum_{n=3}^{\infty} \frac{1}{n!}[{\cal C}_{01}(r)]^n\\
&+\frac{1}{4}\rho_1^2\int d{\bf r}\,h^R_{11}(r) [{\cal C}_{11}(r)]^2
+\frac{1}{2}\rho_1^2 \int d{\bf r}\,g^R_{11}(r) \sum_{n=3}^{\infty} 
\frac{1}{n!}[{\cal C}_{11}(r)]^n\\
&-\frac{1}{4}\rho_1^2\int d{\bf r}\,h^R_b(r)[{\cal C}_{b}(r)]^2
-\frac{1}{2}\rho_1^2\int d{\bf r}\,g^R_{b}(r)
\sum_{n=3}^{\infty} \frac{1}{n!}[{\cal C}_{b}(r)]^n.
\end{split}\end{equation}
The associated EXP approximation for the pair distribution
functions is given by\cite{KRTM1997}
\begin{subequations}
\begin{align}
&g_{01}^{EXP}(r)=g_{01}^R(r)\exp[{\cal C}_{01}(r)]\\
&g_{11}^{EXP}(r)=g_{11}^R(r)\exp[{\cal C}_{11}(r)],\\
&g_b^{EXP}(r)=g_b^R(r)\exp[{\cal C}_b(r)],\\
&g_c^{EXP}(r)=g_{11}^{EXP}(r)-g_b^{EXP}(r).
\end{align}
\end{subequations}

A noticeable property of the above equations describing the structure
and the thermodynamics of a fluid adsorbed in a disordered  matrix
is that the quenched matrix enters solely through its pair
correlation function $h_{00}(r)$. The basic strategy followed in this paper is to
take advantage of this property to study fluid adsorption in a 
model aerogel obtained by  an out-of-equilibrium
aggregation process (see below). The aerogel is then only characterized by its pair distribution function (in
fact, its structure factor obtained by Fourier transform) determined
by computer simulation. At first sight, this may lead
to a conceptual difficulty. Indeed, the original treatment of
quenched-annealed mixtures was restricted to situations in which 
the matrix is  obtained by quenching an equilibrium
distribution \cite{MG1988,G1992,RTS1994}.
 However, as was stressed by Madden \cite{M1995}, there is no such
difficulty and  the RSOZ equations apply in a more general
context, although the diagrammatic structure of the direct correlation functions is
non-standard (indeed, these equations  simply follow from the Legendre
transform that relates the generating functional of the total correlation functions
to the generating functional of the direct correlation functions). 
One can use the OCT approximation scheme as well.

\section{Model systems and numerical methods}
\label{sec-model}

As stressed above, aerogels display very specific structural
properties that can be
expected to have a strong influence on the properties of the adsorbed
fluids. First, they are made of long chains of connected,
almost spherical, colloidal SiO$_2$ particles. Despite the very low
density of the adsorbent, this connectedness
induces  short-range correlations that differ from that generated by
 equilibrium hard-sphere matrices. Secondly, the silica
strands forming the gel are organized into a long-range, fractally
correlated structure whose experimental signature is a pronounced peak
at small wave-vectors in the intensities measured by small-angle
x-ray or neutron scattering. Accordingly, it seems important to
use a modeling of the aerogel structure that reproduces these two
properties.
In the present work, the aerogel is modeled as a connected cluster of
spheres generated
by a diffusion-limited cluster-cluster aggregation (DLCA) process
\cite{M1983,KBJ1983}. This process mimics the mechanisms
involved in the synthesis of real aerogels, and the three-dimensional
off-lattice version used here has been shown to provide gel structures
that correctly account for both the short-range and long-range
structural properties of real silica aerogels \cite{HFPJ93}.

\begin{figure}
\centering
\rotatebox{270}{\scalebox{0.5}{
\includegraphics*{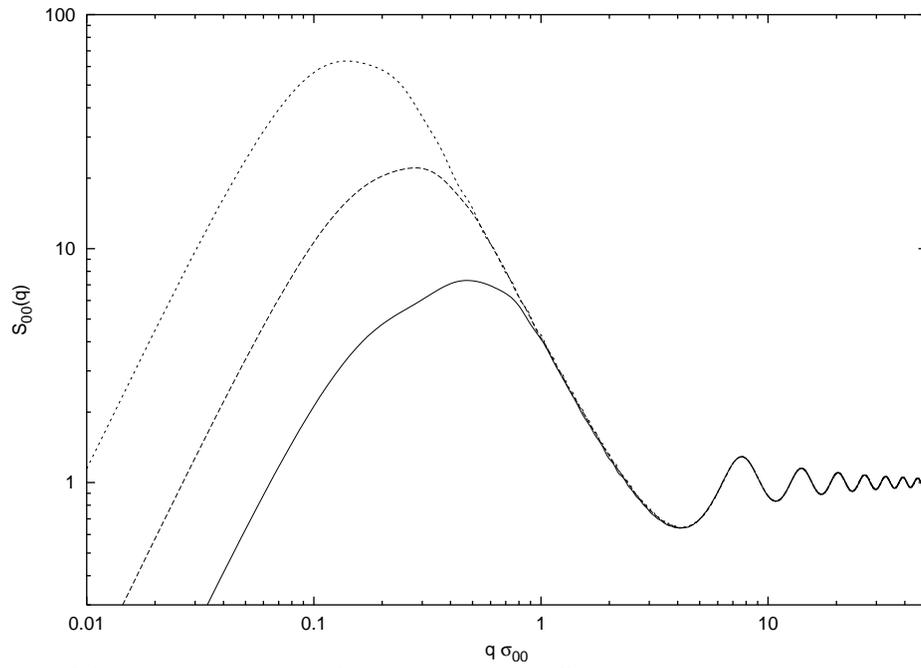}
}}
\caption{\label{aerostruct} DLCA aerogel structure factors at 
three different densities:  $\rho_0=0.025$, 0.05,
0.1 (from top to bottom). These curves represent averages over 40
realizations of the aerogel generated at a given density.}
\end{figure}

For a given matrix density $\rho_0$, the only information about
the matrix that is needed in our calculation is the sample averaged structure
factor $S_{00}(q)$ defined from the averaged pair correlation
function $h_{00}(r)$ as
\begin{equation*}
S_{00}(q)=1+\rho_0
\int_0^{\infty}h_{00}(r)\frac{\sin(q\,r)}{q\,r}\,4\pi\,r^2\,dr.
\end{equation*}
$S_{00}(q)$ has been computed following the lines of the original work
of Hasmy \emph{et al.}\cite{HFPJ93} and the reader is referred to the corresponding
publications for details. The corresponding curves for
the three densities studied in this work ($\rho_0=0.025$, $0.05$,
$0.1$) are reported in Fig. \ref{aerostruct}. Two characteristic 
features of aerogel structures are clearly seen on these data: at
large $q$ ($q\,\sigma_{00}>2\pi$), one finds for all densities
\begin{equation*}
S_{00}(q)\simeq 1+2\frac{\sin (q\,\sigma_{00})}{q\,\sigma_{00}},
\end{equation*}
which corresponds to the fact that each aerogel particle has on average
two neighbors connected to it, whereas at low $q$ ($q\,\sigma_{00}<2$),
a broad peak whose amplitude increases  as density decreases is 
found, that originates from the long-range, fractal correlations of
the particle strands forming the material.

The matrix-fluid and fluid-fluid interactions are split into a hard-sphere 
part and a Lennard-Jones 12-6 tail following Weeks, Chandler and
Andersen \cite{WCA1971}, i.e.,  
$u_{\alpha \beta}(r)= -\varepsilon_{\alpha \beta}$
for $\sigma_{\alpha \beta}<r<2^{1/6}\sigma_{\alpha \beta}$ and 
$u_{\alpha \beta}(r)=4\varepsilon_{\alpha \beta}[(\sigma_{\alpha
\beta}/r)^{12}-(\sigma_{\alpha \beta}/r)^{6}]$ for $r>2^{1/6}
\sigma_{\alpha\beta}$.
The use of hard-sphere cores for all the repulsive potentials is
chosen to avoid difficulties associated with calculating the free
energy of fluids with soft-core repulsive forces. Moreover, the 
Lennard-Jones tails are truncated at $r/\sigma_{\alpha \beta}=2.5$.

In order to make contact with previous work \cite{KRTM1997}, we have only
considered systems in which all particles have the same diameter,
i.e., $\sigma_{00}=\sigma_{01}=\sigma_{11}=\sigma$. It has to be
stressed that this choice is rather unrealistic with respect to the actual
experimental situation, since the silica particles forming the aerogels
are always much larger than the adsorbate (typically, by a factor of
ten). However, the model captures most of the specific nature of the
correlated disorder induced by an aerogel and this is expected to be
sufficient to give an account of the generic
experimental behavior. Like in previous studies, we introduce the
interaction ratio $y=\varepsilon_{01}/\varepsilon_{11}$, and, in the
following, $\sigma$ will be used as the length unit,
$\varepsilon_{11}$ as the energy unit and $\varepsilon_{11}/k_B$ as
the temperature unit.

In all calculations, the correlation functions of the reference system
were obtained by solving the RSOZ equations in the Percus-Yevick 
approximation and the corresponding 
pressure and chemical potential were obtained by integration of the compressibility equation of
state, Eq. (13). This is an additional
approximation in the ORPA+B$_2$/EXP framework. However, we know from
previous studies \cite{R1999} that the PY theory
gives a good description of the reference-system properties when
matrix and fluid particles have the same size. However, because these
previous studies have dealt only with matrices quenched from an equilibrium
distribution, we have tried to obtain another assessment of this
approximation for the present situation of an ``out-of-equilibrium'' matrix
by computing 
the Henry's constant $K_H$ of the reference system. For a hard-sphere
system, $K_H$ is simply the fraction of the total volume of the matrix
that is accessible to the center of a single adsorbate molecule.
Thus, we have first estimated $K_H$ by direct Monte-Carlo integration for
various realizations of the model aerogels at the three working
densities. Then, we have computed the theoretical value by using the
(scaled-particle-theory) charging process expression (remembering that
$K_H=\exp[-\beta\mu_{ex}^R(\rho_1=0)]$),
\begin{equation*}
K_H=\exp\left(-\rho_0\int_0^{\sigma} G_{01}(r;r;\rho_1=0) 
4\pi r^2 dr\right),
\end{equation*}
where $G_{01}(r;r;\rho_1=0)$ is the contact value at zero adsorbate
density of the fluid-matrix pair distribution function when the
corresponding hard-core diameter equals $r$ \cite{RFL1959}. The value
of $G_{01}(r;r;\rho_1=0)$ is determined in the Percus-Yevick
approximation for various values of $r$ between 0 and $\sigma$ and the
integral is estimated as a discrete sum. Note that a similar
calculation has already been proposed by Ford {\it et al.}\cite{FTG1995} and
appeared to be quite successful. The results of both
estimations are reported in table \ref{henry} and show a very good
agreement, thereby confirming the accuracy of the PY description of
the reference system.  

In order to solve the RSOZ equations, we used Gillan's method
\cite{G1979} with a grid size of $0.02\sigma$ and 16384 points in most
of the calculations. Once the correlation functions were obtained, 
we deduced the renormalized potentials and computed the free energy to
the chosen approximation. In order to calculate the phase diagram of
the fluid inside the matrix, series of adsorption isotherms $\mu$ vs
$\rho_1$ were generated. When the curve had a loop, similar to those 
seen in the van der Waals theory of equilibrium fluids, the densities
$\rho_1'$ and $\rho_1''$ of the two coexisting phases were determined
by solving the system of equations
\begin{equation*}
P(\rho_{1}')=P(\rho_{1}''), \qquad\qquad
\mu(\rho_{1}')=\mu(\rho_{1}''),
\end{equation*}
where  $P=-\left.\partial \overline{A}/\partial V\right|_{T,\rho_0}$
is the grand potential density (or thermodynamic pressure of the fluid
\cite{RTS1994}).

\begin{figure}
\centering
\rotatebox{270}{\scalebox{0.6}{
\includegraphics*[0cm,0cm][21cm,21cm]{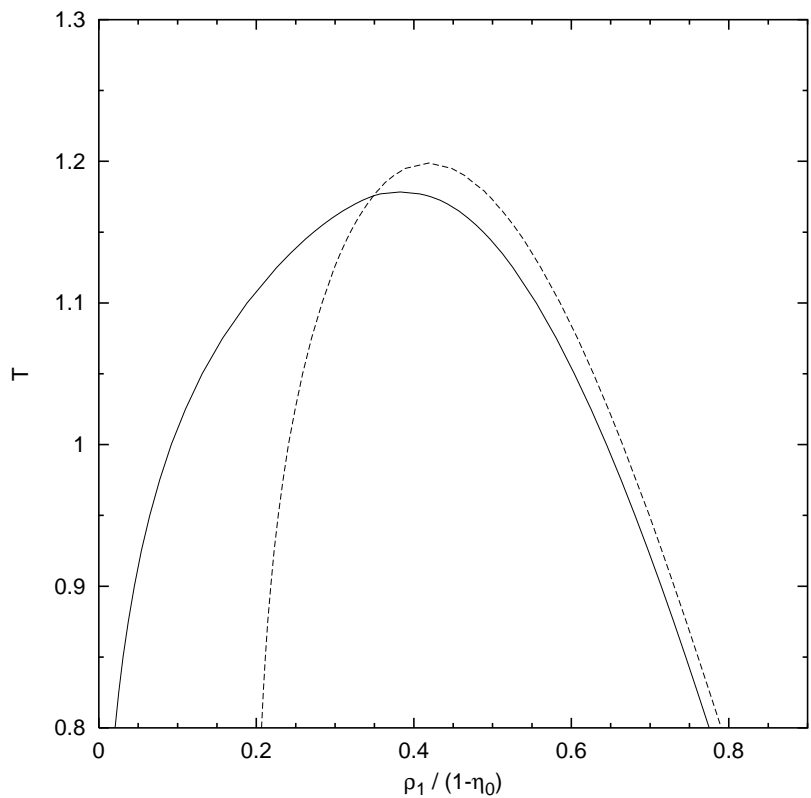}
}}
\caption{\label{compapprox} Comparison of the coexistence envelopes for a
fluid confined in an aerogel matrix ($\rho_0=0.05$, $y=1.5$) computed
within the MSA (solid line) and the ORPA+B2 (dashed line).}
\end{figure}

\begin{figure}
\centering
\rotatebox{270}{\scalebox{0.6}{
\includegraphics*[0cm,0cm][21cm,21cm]{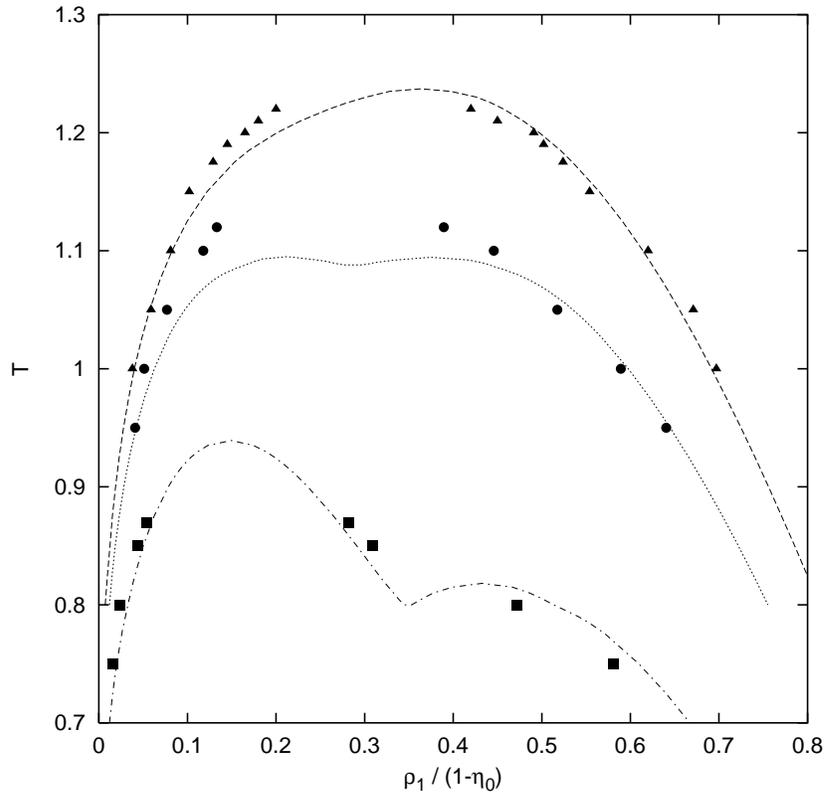}
}}
\caption{\label{levesque} Comparison of the coexistence envelopes for a
fluid confined in purely repulsive ($y=0$) equilibrium hard-sphere
matrices ($\rho_0=0 \text{ (bulk)}$, $0.046$ and $0.15$ from top to bottom)
obtained by computer simulations (symbols) \protect\cite{ALW1999} and within
the ORPA+B2 (lines) \protect\cite{KRTM1997}.} 
\end{figure}

In Ref.\cite{KRTM1997}, it was shown that there is a great sensitivity of
the calculated phase diagrams to the approximations used in the
theory. This is the case in the present calculation as well, as can be
seen in Fig. \ref{compapprox}, where the MSA and ORPA+B2 phase
diagrams are compared for typical values of the parameters. Note that
the fluid density has been normalized by the porosity of the matrix
that is equal to $1-\eta_0$, where $\eta_0=\pi\rho_0\sigma^3/6$ is the corresponding packing
fraction. In the following, we will concentrate on
the results of the ORPA+B2/EXP approximation. There are two
motivations to this choice. The first one is of fundamental
nature. Experiments show unambiguously that it is the low-density part
of the phase diagram that is the most strongly affected by the
presence of an aerogel. So, it is sensible to address the problem with a
theory well adapted to this low-density regime. For bulk fluids, it
 is well known that, at
variance with the lower-order approximations ORPA or MSA, the ORPA+B2
theory predicts the free energy with the same level of accuracy in the
low-density gas and the high-density liquid phases and that the EXP
approximation gives the exact low- density limit for the
pair distribution functions \cite{AC1972}. This is not true
anymore when one considers quenched-annealed system (for instance, the
ORPA+B2 free energy does not yield in general the exact second virial
coefficient for the fluid inside the matrix, because the term of order
$\rho_1^2$ contains contributions from all orders in $\rho_0$), but one still 
expects that the ORPA+B2/EXP theory will remain
accurate in the case of a low-density gas adsorbed in a low-density
matrix. The second reason is
more pragmatic and comes from the comparison with  the simulation results of {\'A}lvarez
{\it et al.} \cite{ALW1999} for hard-sphere matrices 
of low and moderate densities. In this case,  the ORPA+B2
phase diagrams turn out to give the best agreement with the simulation
results. This is shown in Fig. \ref{levesque}, where the simulation
results tabulated in Ref.\cite{ALW1999} are compared with the theoretical
curves of Ref.\cite{KRTM1997}: the agreement is reasonably good (much
better indeed that with MSA).

\section{Results}
\label{sec-results}

We first present the results for the structure of the adsorbed fluid,
since it shows remarkable features originating in the specific
properties of the aerogel, in particular its
connectedness.

\begin{figure}
\centering
\rotatebox{270}{\scalebox{0.63}{\includegraphics*{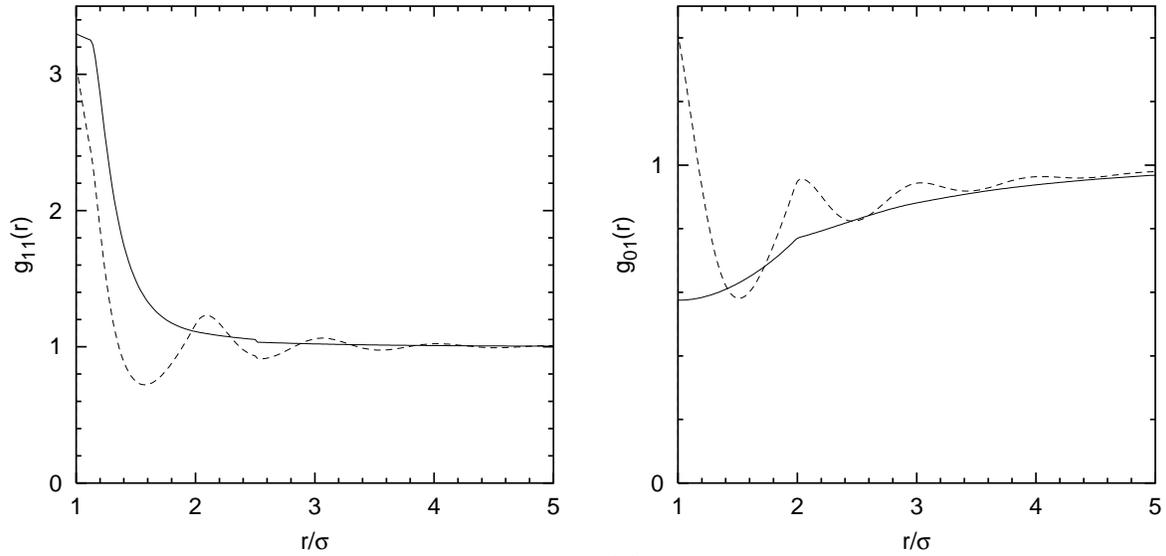}}}
\caption{\label{corrhole} Fluid-fluid distribution function $g_{11}(r)$
and matrix-fluid  distribution function $g_{01}(r)$ calculated with
the EXP approximation in an aerogel matrix with $\rho_0=0.05$ and
$y=0$ at $T=0.9$. Solid line: $\rho_1=0.02$; dashed line:
$\rho_1=0.70$.} 
\end{figure}

The first effect of the aerogel structure shows up already in the case of
purely repulsive matrix-fluid interactions ($y=0$). This is
illustrated in Fig. \ref{corrhole}, where the matrix-fluid
correlation function $g_{01}(r)$ at gas density ($\rho_1=0.02$) displays a very clear
depletion near contact. This ``correlation hole'' comes from the
presence of a region surrounding each matrix particle from which the adsorbate molecules
are excluded. This region disappears when density is increased, as can
be seen from the $g_{01}(r)$ at liquid density ($\rho_1=0.70$). This correlation hole,
as well as the accompanying cusp at $r/\sigma=2$, is a well-known
feature of molecular systems made of tangentially jointed hard-spheres, like our model aerogel. For instance, it has been found in
fluids of freely-jointed hard-sphere chains, by both computer
simulation and theory \cite{corrhole}. It is a simple 
steric effect due to the presence of the neighboring beads that
hinders the approach of a gas molecule.

The second effect, expected to manifest itself at low enough
temperatures when attractive matrix-fluid interactions are turned on,
is a much stronger binding of the adsorbate particles to the solid
matrix than in the previously investigated models. Indeed, on one
hand, two tangent aerogel beads form adsorption sites with potential
energy equal or lower than $-2\varepsilon_{01}$ (the direct contribution
of the two  beads being still lowered by contributions from
their neighboring beads); on the other hand, the aerogel structure
presents large voids \cite{HFPJ93,PP1999} in which the adsorption potential
energy due to the matrix is zero. This leads to a highly inhomogeneous
potential energy field and thus to strong preferential adsorption on
the aerogel strands. The situation is very different in the case of an
equilibrium hard-sphere matrix at a similar density, as shown by the
following simple calculation. For such a matrix at a density
$\rho_0=0.05$, the average distance between two spheres is roughly
$d=(\rho_0)^{-1/3}\simeq 2.7$. If an adsorbate particle is equidistant
from two matrix particles, it experiences from these particles a
potential energy roughly equal to
$2u_{10}(r=d/2)=-1.075 \ \varepsilon_{01}$, whereas if it is adsorbed on a
matrix particle, it experiences a potential energy simply equal to
$-\varepsilon_{01}$. The situation is thus far less contrasted than in an
aerogel and one
expects a relatively weak binding of the adsorbate on the matrix. 

\begin{figure}
\centering
\rotatebox{270}{\scalebox{0.63}{
\includegraphics*{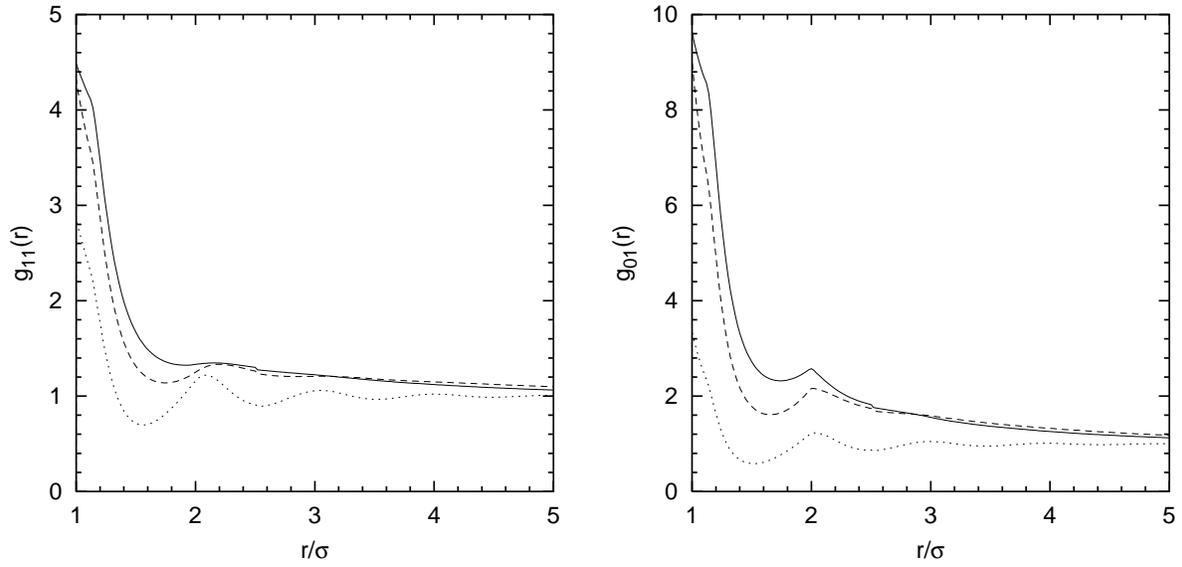}
}}
\caption{\label{corrads} Fluid-fluid distribution function $g_{11}(r)$
and matrix-fluid distribution function $g_{01}(r)$ calculated with
the EXP approximation in an aerogel matrix with $\rho_0=0.05$ and
$y=1.5$ at $T=0.9$. Solid line: $\rho_1=0.02$; dashed
line: $\rho_1=0.2$; dotted line: $\rho_1=0.72$.} 
\end{figure}

\begin{figure}
\centering
\rotatebox{270}{\scalebox{0.63}{
\includegraphics*{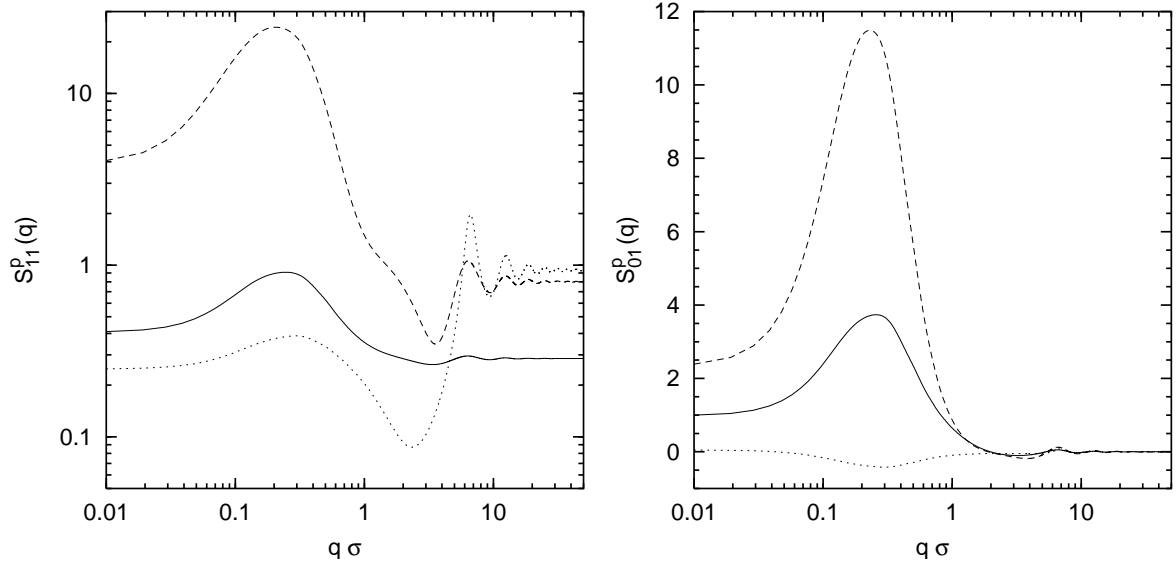}
}}
\caption{\label{structfact} Partial fluid-fluid structure factor
$S^p_{11}(q)$ and matrix-fluid structure factor $S^p_{01}(q)$
calculated within the EXP approximation in an aerogel matrix with
$\rho_0=0.05$ and $y=1.5$ at $T=0.9$. Solid line:
$\rho_1=0.02$; dashed line: $\rho_1=0.2$; dotted line:
$\rho_1=0.72$. Note that the oscillations of $S^p_{01}(q)$ around
zero at large $q$  prevent us from using a logarithmic scale for $S^p_{01}(q)$.}
\end{figure}

This second effect should result in highly inhomogeneous distributions
of the fluid particles and is thus difficult to detect from averaged
quantities like pair distribution
functions. We find nevertheless a few indications of its 
influence on the adsorbate structure. One of them can be seen in
Fig. \ref{corrads}. It is indeed quite direct to see from the values
of the pair distribution functions near contact that, at the two vapor
densities considered ($\rho_1=0.02$ and $\rho_1=0.2$), there are
roughly twice more adsorbate particles around a matrix particle than
around a given adsorbate particle. This is compatible with the picture
of the formation of a strongly bound monolayer of particles adsorbed
on the aerogel strands. It agrees with the statement by
Chan \emph{et al.} that the ``gas'' phase in very dilute aerogels actually consists of
vapor plus a liquid film that coats  the random silica strands
\cite{WC1990}. This situation differs strongly from that in
equilibrium hard-sphere matrices, where the prevailing adsorption  mechanism is
expected to be the formation of
liquids droplets in the denser regions of the matrix. In this latter case, the environments of the matrix and fluid
particles are not too different. This is precisely what is found in
Ref.\cite{KRTM1997} (see Fig.~13 of this paper and the related
discussion). Another indication of strong  adsorption
on the aerogel strands is the presence of structural 
correlations typical of the aerogel structure in the adsorbed low-density
phase. Indeed, if the adsorbate particles coat the aerogel
strands, it is expected that their distribution will retain part of
the spatial arrangement of the substrate, in particular the
long-range correlations that show up in the form of a maximum at
small wave vectors. To test this idea, we have calculated the partial
structure factors of the system defined as \cite{HMcD1976}
\begin{align*}
S_{11}^p(q)=&\frac{\rho_1}{\rho_0+\rho_1}(1+
\rho_1\,\hat{h}_{11}(q)),\\
S_{01}^p(q)=&\frac{\rho_0}{\rho_0+\rho_1}\,
\rho_1\,\hat{h}_{01}(q).
\end{align*}
Results are shown in Fig. \ref{structfact}, where it is clearly seen
that a broad correlation peak builds up at small $q$ ($q\sigma \simeq  0.2$) when increasing
the vapor density, at a position in good agreement with the position of
the corresponding peak in the aerogel structure factor (compare with
Fig. 1). This can again
be interpreted as evidence for the progressive formation of a thin
layer of adsorbed particles coating the gel strands.

These characteristic  features of adsorbed fluid phases in aerogels
are important at low adsorbate density only. Indeed, as can be seen in Figs.
\ref{corrhole} to \ref{structfact}, they are almost completely absent
at liquid densities, for which  the physics of the system is dominated by the hard-core
interaction between adsorbate particles so that the structure of the
adsorbent becomes of marginal importance. Thus, these observations
stress \emph{a posteriori} the need to use a theoretical framework
well adapted to the description of the low-density regime, hence the
use of the ORPA+B2/EXP theory in this work. But they also show that
the case under study is an especially difficult one in the context of
the OCT: indeed, of the two effects discussed
above, the first one, that is purely steric in origin, competes with the second
one, that is only due to the attractive matrix-fluid interactions.

We now turn to the presentation of the capillary phase diagrams.
 We only report the results corresponding to the case where both
fluid-fluid and matrix-fluid attractive interactions are present,
which is the case of experimental relevance.  

\begin{figure}
\centering
\rotatebox{270}{\scalebox{0.6}{
\includegraphics*[0cm,0cm][21cm,21cm]{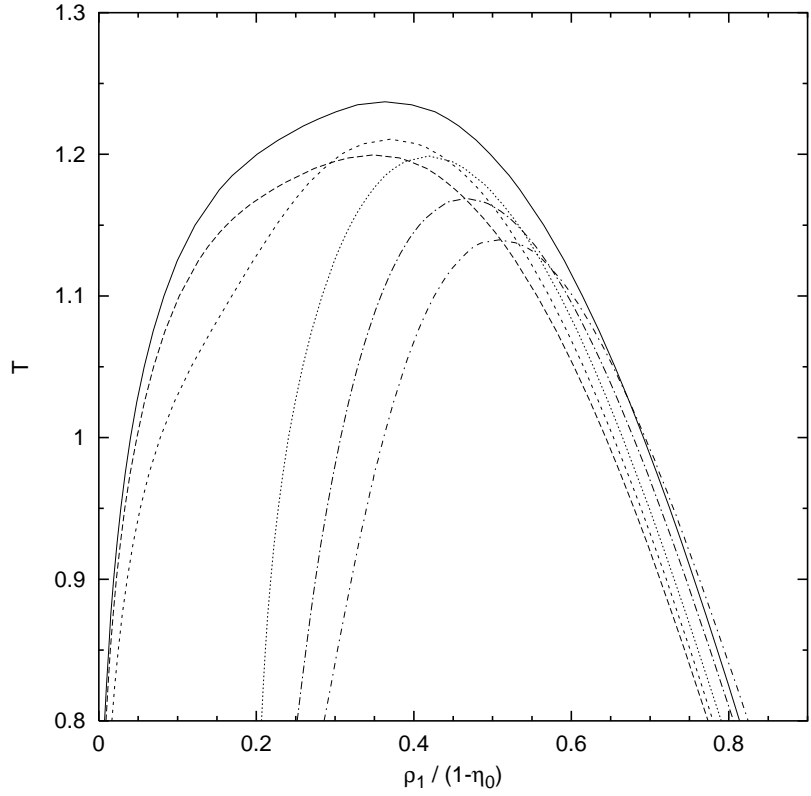}
}}
\caption{\label{ynonzero} ORPA+B2 predictions for the coexistence
envelopes of the fluid phases in an aerogel matrix with density $\rho_0=0.05$.  From
left to right: $y=0.5$, 1.0, 1.5, 2.0, 2.5. The solid line is for the
bulk fluid.}
\end{figure}

Fig.\ref{ynonzero} shows coexistence envelopes of the fluid phases
obtained from the ORPA+B2 approximation as one varies the  ratio $y$
for a typical aerogel density
$\rho_0=0.05$ corresponding to a porosity $1-\eta_0=97.4$\%. It is
seen that the variation of the capillary critical temperature with $y$ is
non-monotonic: $T_{cc}$ first increases and then decreases as one
increases $y$, the maximum value being found around $y=1$. On the
other hand, the critical density 
increases monotonically and becomes larger than the bulk value for
$y=1$ (this not due to the normalization by $1-\eta_0$). The latter behavior agrees with the experimental observation
that $\rho_{1c}$ is displaced towards the liquid phase when the
interaction with the porous material is sufficiently attractive
\cite{WC1990}. Despite the low density of the aerogel, there is also a
significant narrowing of the coexistence curves compared to the bulk
one when $y>1$, although this effect is still smaller than that
observed in the experiments (see below) \cite{WC1990,Z1996}.

Fig. \ref{varydens} shows the effect of varying the aerogel density
at a fixed value of the interaction ratio $y=1.5$. It is found that
the normalized critical density increases only slightly with the
matrix density and that the width of the coexistence curve does not
vary very much. However, for $y=1.5$ and $\rho_0=0.025$, a 
``precondensation'' transition is obtained, similar to what has
already been found
in various models \cite{R1999,KRTM1997} (see also Maritan {\it et al.}\cite{MS1991}). A surprising
result seen in this figure is the non-monotonic variation
of the capillary critical temperature that first increases with $\rho_0$ and then
decreases, leading to the somewhat unexpected result that for a low
enough matrix density $T_{cc}$ is larger than its bulk value. The
physical origin, if any, of this behavior is unclear to us, but there is \emph{a priori} no fundamental
reason forbidding such a behavior. A possible explanation could be  the very weak confinement effect due to the dilute
gel. Indeed, it is well known, already at the mean-field level, that
confinement induces a lowering of the critical temperature of a
fluid. In particular, this is what is found in all cases of simple pore
geometries \cite{FLC1994}. But aerogels are rather special in this
respect: because of their very open structure, they have no
real pores with well defined walls. The comparison with the case of
equilibrium hard-sphere matrices gives an insight on the potential
relevance of this fact. Indeed, in the latter case, it has been found
that adsorption partly counterbalances the influence of confinement,
so that the lowering of the critical temperatures is less pronounced
in the case of attractive fluid-matrix interactions than in the
completely repulsive case. A possible origin of our finding for
aerogels could thus be that the effect of the attractive matrix-fluid 
interaction now dominates the very weak influence of confinement.

The remaining question is of course to see how our results based on a
realistic description of the structural correlations in an aerogel
compare with experiments and with previous results obtained with the
less realistic equilibrium hard-sphere model.
 
As already stated in the introduction, the most significant
experimental observation is a marked narrowing of the liquid-vapor coexistence
curve relative to that in the bulk, a narrowing that is accompanied by a slight
displacement of the critical point to a lower temperature and a higher
density. Chan \emph{et al.}\cite{WC1990} have proposed to quantify this narrowing
by fitting the top of the coexistence curves for ${}^4$He and N$_2$
adsorbed in 95\% aerogels with the critical expression
\begin{equation*}
\frac{\rho_l-\rho_v}{\rho_{1c}}=B\,\left[\frac{T_{cc}-T}{T_{cc}} \right]^\beta,
\end{equation*}
where $\rho_v$ and $\rho_l$  are, respectively, the vapor and 
liquid  densities of the adsorbed fluid, and they have found
that the amplitude factors $B$ are respectively about 14 and 2.6 times
smaller than the bulk values for these two fluids  (in
both cases, the value of the exponent $\beta$ is consistent with that
found for the bulk fluids). These
values, which seem indicative of a large reduction of the phase
coexistence envelopes, have nevertheless to be taken with
care. Indeed, in the case of ${}^4$He, the previous result has been
obtained from data in the temperature range 5.15--5.17 K, $T_{cc}$ being
found equal to 5.167 K. Further experiments in a wider temperature
domain (2.5--5 K) have shown that the reduction of the phase envelope
is much more pronounced near the
critical point than well below the critical temperature
\cite{CHAN}. For instance, when the temperature is roughly
$0.64\,T_c$, the ratio of the widths of the bulk and
adsorbed fluid phase diagrams is not larger than about 2.2.

As can be seen in Figs. \ref{ynonzero} and \ref{varydens}, for
values of $y$ that are large enough, one finds that our model gives
results in good qualitative agreement with the picture emerging from
experiments: the critical density is slightly shifted to values higher than in
the bulk and, except for the lowest aerogel density, the critical
temperature is found to decrease, albeit moderately. A significant narrowing of the
coexistence curves is also found. For instance, with $\rho_0=0.05$ and
$y=1.5$, the ratio of the widths of the bulk and adsorbed fluid phase
diagrams is about 1.4 at $T=0.8\simeq0.65\,T_c$ and it
increases with $y$. (However, our present approach is unable to capture
the behavior seen in the ${}^4$He experiment in the close vicinity of $T_{cc}$.)

Figs. \ref{compar} (a) and \ref{compar} (b) show a comparison between the coexistence curves
obtained for a  fluid adsorbed in model aerogels and equilibrium
hard-sphere matrices. It is seen that, at given matrix density $\rho_0$, and
ratio of interactions $y$, the coexistence curves in the case of the
model aerogel are
always narrower, and the critical temperatures always higher, than in the
equilibrium hard-sphere case. The effect is quite significant for $y=1.5$, where the
ratio of the widths of the bulk and adsorbed fluid phase envelopes is
about 90 \% at $T=0.8\simeq0.65\,T_c$ for hard-sphere
matrices, compared to about 70 \% for aerogels. Combined with the fact
that the width of the coexistence curves varies only weakly with the
aerogel density, we thus find that using a more realistic model of the
adsorbent makes it possible to obtain a behavior that  more closely resembles
the experimental one, i.e., a substantial narrowing of the phase
envelope associated to weak changes in the critical temperature and
density. (It should also be stressed that for the small matrix
densities used here, the hard-sphere matrix is far from being a
connected medium.) We note however that this improvement of the situation
compared to experiments is limited to a certain range of parameters,
since the differences between the two types of matrices become rather
modest for larger values of $y$ ($y>2$), for which it seems that the 
strength of the attractive matrix-fluid potential limits the relevance
of the detailed structure of the adsorbent.

\section{Conclusion}

In this paper, we have presented a theoretical study of the phase
diagram and the structure of simple fluids adsorbed in high-porosity
aerogels. For these systems, the perturbation induced by the
adsorbent is {\it a priori} small, and one expects the
integral-equation approach combined with the replica formalism to
provide a reasonable description. The novelty of the present work is
that a realistic description of the aerogel structure has been
incorporated via the use of the static structure factor of model
aerogels built from an off-lattice diffusion limited cluster-cluster
aggregation process. Unlike cruder models based on quenched
configurations of equilibrium hard or penetrable spheres, the
correlations between matrix particles at the low density studied
(corresponding to porosities between $90\%$ and $97.5\%$) keep track of
the short-range connectedness of the aerogel and of the long-range
fractal behavior. The predictions of the theory are in qualitative
agreement with the experimental results, showing a significant
narrowing of the gas-liquid coexistence curve associated with weak
changes in the critical temperature and density. In addition, the
influence of the aerogel structure is shown to be important at low
fluid densities. Improving the present approach is still needed in two
directions: first, to provide a better description of the critical
region, but this represents a very challenging task for systems like
fluids adsorbed in disordered porous media that are akin to random
fields models; secondly, to go one step further toward a realistic
modeling of the adsorbent by allowing the beads composing the silica
strands of the aerogel to be much larger (typically, by a factor of ten)
than the fluid particles. The latter improvement of the description
requires both the development of more appropriate closures of the
Ornstein-Zernike equations  for mixtures of particles with very
different sizes and the introduction of a more realistic potential
between the matrix and the fluid particles (using for instance the
composite-sphere potential proposed in Ref.\cite{KM1991}).

\vspace{2cm}

The Laboratoire de Physique Th{\'e}orique des Liquides is UMR 7600
of the CNRS. We thank R. Jullien for providing us with his DCLA
algorithm to build the model aerogels and M. Chan for communicating
unpublished experimental results.

\begin{table}
\begin{tabular}{cccc}
$\rho_0$ & MC & PY & EHS\\\hline
0.025  	& 	0.934 	& 	0.927 	&	0.898	\\
0.05  	& 	0.868 	& 	0.856 	&	0.802	\\
0.1   	&	0.739 	& 	0.718	&	0.629
\end{tabular}
\vspace{1cm}

\caption[]{\label{henry} Henry's constant of a hard-sphere fluid
adsorbed in DLCA aerogels, calculated by direct Monte-Carlo
integration (MC) and within the Percus-Yevick approximation (PY) through a
scaled-particle-theory charging process. The fluid and matrix
particles have the same diameter. For comparison, the values for
hard-sphere matrices (EHS)  calculated as in Ref. \cite{KRTM1997} are also
given.}
\end{table}

\begin{figure}
\centering
\rotatebox{270}{\scalebox{0.6}{
\includegraphics*[0cm,0cm][21cm,21cm]{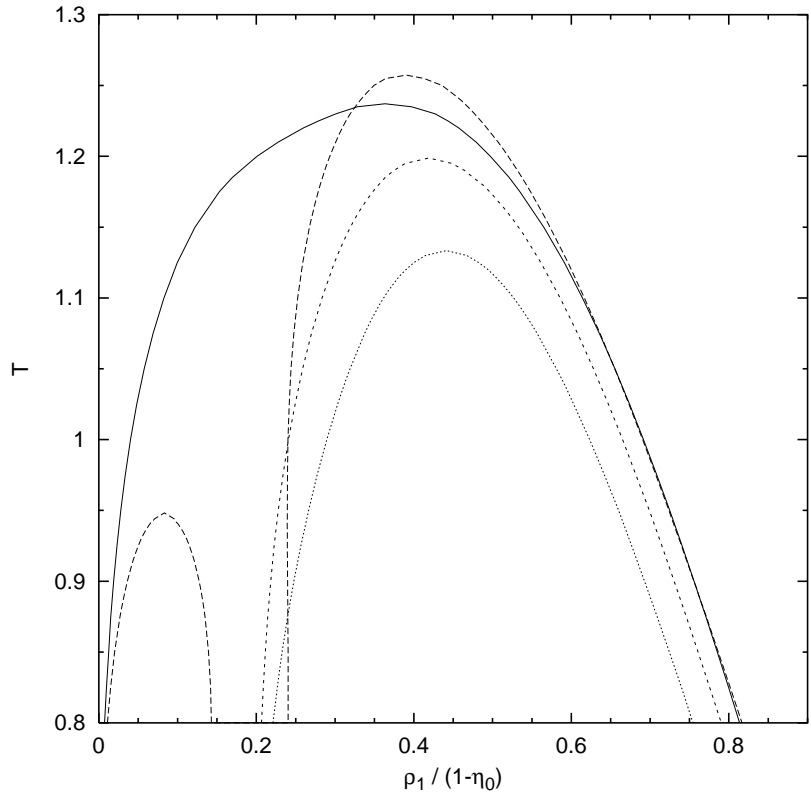}
}}
\caption{\label{varydens} ORPA+B2 predictions for the coexistence
envelopes of the fluid phases in an aerogel matrix with interaction ratio $y=1.5$. From
top to bottom, $\rho_0=0.025$, 0.05, 0.1. The solid line is for the
bulk fluid.}
\end{figure}

\begin{figure}
\centering
\rotatebox{270}{\scalebox{0.63}{
\includegraphics*{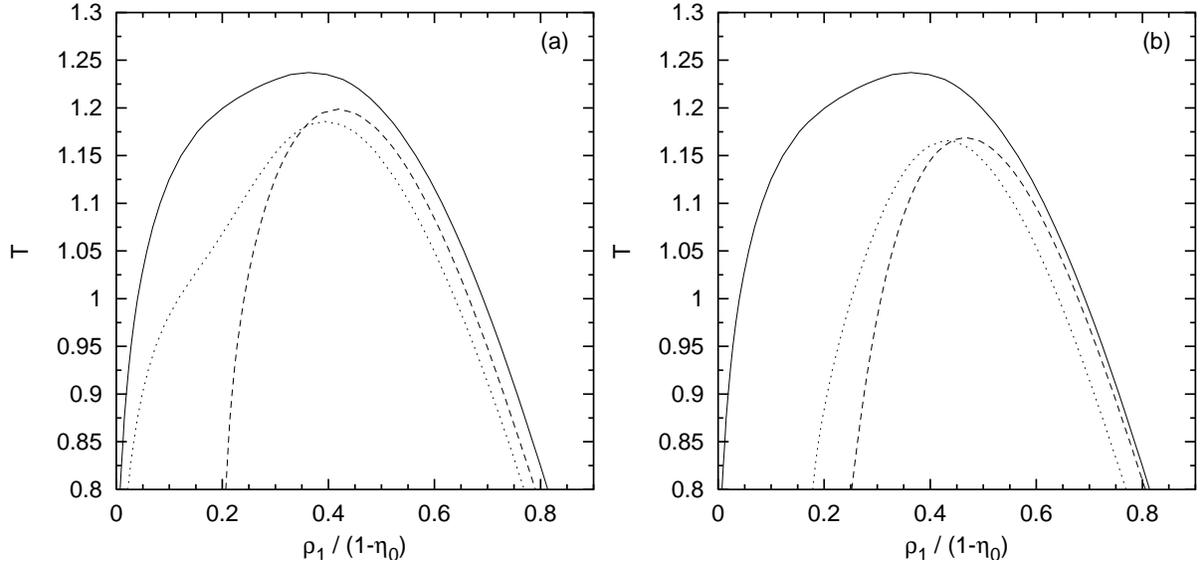}
}}
\caption{\label{compar} Comparison of the gas-liquid coexistence envelopes
predicted by ORPA+B2 in an equilibrium hard-sphere matrix and a DLCA
aerogel at the same matrix density $\rho_0=0.05$. (a) $y=1.5$. 
Dotted line: hard-sphere matrix; dashed line:
aerogel. (b) $y=2.$ Dotted line: hard-sphere matrix; dashed line:
aerogel.
The solid line is for the bulk
fluid.}
\end{figure}
\end{document}